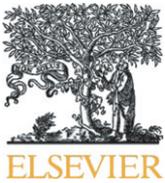
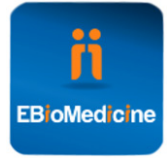
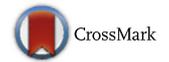

Research Paper

# ADAM30 Downregulates APP-Linked Defects Through Cathepsin D Activation in Alzheimer's Disease

Florent Letronne [a,b,c,1], Geoffroy Laumet [a,b,c,1], Anne-Marie Ayral [a,b,c,1], Julien Chapuis [a,b,c,1], Florie Demiautte [a,b,c,1], Mathias Laga [d,e], Michel E. Vandenberghe [f,g], Nicolas Malmanche [a,b,c], Florence Leroux [b,c,h], Fanny Eysert [a,b,c], Yoann Sottejeau [a,b,c], Linda Chami [i,j], Amandine Flaig [a,b,c], Charlotte Bauer [i,j], Pierre Dourlen [a,b,c], Marie Lesaffre [k], Charlotte Delay [a,b,c], Ludovic Huot [b,c,l], Julie Dumont [b,c,h], Elisabeth Werkmeister [b], Franck Lafont [b], Tiago Mendes [a,b,c], Franck Hansmannel [m,n], Bart Dermaut [a,b,c], Benoit Deprez [b,c,h], Anne-Sophie Hérard [f,g], Marc Dhenain [f,g], Nicolas Souedet [f,g], Florence Pasquier [o,p], David Tulasne [k], Claudine Berr [q], Jean-Jacques Hauw [r], Yves Lemoine [b,c,l], Philippe Amouyel [a,b,c,p], David Mann [s], Rebecca Déprez [b,c,h], Frédéric Checler [i,j], David Hot [b,c,l], Thierry Delzescaux [f,g], Kris Gevaert [d,e], Jean-Charles Lambert [a,b,c,*]

[a] INSERM, U1167, Laboratoire d'Excellence Distalz, F59000 Lille, France
[b] Institut Pasteur de Lille, F59000 Lille, France
[c] Univ. Lille, F59000 Lille, France
[d] Department of Medical Protein Research, VIB, Ghent, Belgium
[e] Department of Biochemistry, Ghent University, Ghent, Belgium
[f] CEA, DSV, I2BM, MIRCen, Fontenay aux Roses, France
[g] CNRS, UMR 9199, Fontenay aux Roses, France
[h] INSERM U1177, Drugs and Molecules for Living Systems, F5900 Lille, France
[i] Institut de Pharmacologie Moléculaire et Cellulaire, UMR 7275 CNRS, Laboratoire d'Excellence Distalz, Nice, France
[j] Université de Nice-Sophia-Antipolis, Valbonne, France
[k] Univ. Lille, CNRS, Institut Pasteur de Lille, UMR 8161 – M3T – Mechanisms of Tumorigenesis and Targeted Therapies, F-59000 Lille, France
[l] Center for Infection and Immunity of Lille, CNRS UMR 8204, INSERM 1019, Lille, France
[m] INSERM, U954, Vandoeuvre-lès-Nancy, France
[n] Department of Hepato-Gastroenterology, University Hospital of Nancy, Université Henri Poincaré 1, Vandoeuvre-lès-Nancy, France
[o] Univ. Lille, Inserm, U1171, - Degenerative & Vascular Cognitive Disorders, Laboratoire d'Excellence Distalz, F-59000 Lille, France
[p] CHR&U, Lille, France
[q] INSERM, U1061, Université de Montpellier I, Hôpital La Colombière, Montpellier, France
[r] APHP-Raymond Escourolle Neuropathology Laboratory, la salpétrière Hospital, Paris, France
[s] Institute of Brain, Behaviour and Mental Health, University of Manchester, Salford Royal Hospital, Salford, UK

## ABSTRACT

Although several ADAMs (A disintegrin-like and metalloproteases) have been shown to contribute to the amyloid precursor protein (APP) metabolism, the full spectrum of metalloproteases involved in this metabolism remains to be established. Transcriptomic analyses centred on metalloprotease genes unraveled a 50% decrease in *ADAM30* expression that inversely correlates with amyloid load in Alzheimer's disease brains. Accordingly, in vitro down- or up-regulation of *ADAM30* expression triggered an increase/decrease in Aβ peptides levels whereas expression of a biologically inactive ADAM30 (ADAM30$^{mut}$) did not affect Aβ secretion. Proteomics/cell-based experiments showed that ADAM30-dependent regulation of APP metabolism required both cathepsin D (CTSD) activation and APP sorting to lysosomes. Accordingly, in Alzheimer-like transgenic mice, neuronal ADAM30 over-expression lowered Aβ42 secretion in neuron primary cultures, soluble Aβ42 and amyloid plaque load levels in the brain and concomitantly enhanced CTSD activity and finally rescued long term potentiation alterations. Our data thus indicate that lowering ADAM30 expression may favor Aβ production, thereby contributing to Alzheimer's disease development.

## 1. Introduction

Alzheimer's disease is a complex, multifactorial, neurodegenerative disease. It is the leading cause of dementia in elderly people. The main pathologic features of Alzheimer's disease are neurofibrillary tangles and senile plaque formation in the brain. The latter is caused by the progressive deposition of mainly 39- to 43-amino acid amyloid β (Aβ) peptides generated by proteolytic cleavage of the amyloid precursor protein (APP). The systematic observation of changes in APP metabolism in monogenic forms of Alzheimer's disease suggested that the Aβ/APP pathway is at the heart of the disease (Hardy and Selkoe, 2002). Even though the key role of APP processing in the etiology of Alzheimer's disease has been challenged in recent years, recent genetic and GWAS studies of sporadic forms of Alzheimer's disease seem to support the importance of dysfunctional APP metabolism and Aβ peptide production/degradation in the physiopathology of Alzheimer's disease (Lambert and Amouyel, 2011; Jonsson et al., 2012; Tian et al., 2013; Young et al., 2015).

The two major Aβ peptide species (Aβx-40 and Aβx-42) are produced by the sequential endoproteolysis of APP by β-secretase and γ-secretase complexes. APP can also undergo non-amyloidogenic cleavage by α-secretase within the Aβ sequence, which thereby precludes Aβ generation. The various enzymes and protein complexes involved in these secretase activities are increasingly well characterized: β-site APP cleaving enzyme 1 (BACE1) accounts for almost all the β-secretase activity, whereas a complex that includes presenilin 1 or 2 is responsible for the γ-secretase activity (De Strooper, 2003). More recently, an additional matrixine (MMP-MT5) has been shown to contribute to APP processing by acting upstream of the β-site (Willem et al., 2015; Baranger et al., 2016). Finally other N-terminal truncated Aβ peptide species can be generated by other proteases besides BACE1 (Wang et al., 2006; Schönherr et al., 2016).

In contrast, α-secretase activity is less well characterized even if several research groups have suggested that ADAM10 and ADAM17 are the major enzymes responsible for constitutive and regulated α-secretases-mediated pathways in the brain (Lammich et al., 1999; Kuhn et al., 2010). In addition to these direct cleavages taking place on APP, several additional proteins are likely to modulate APP levels by interfering with secretase activity, APP trafficking and/or APP degradation (Vincent and Checler, 2012). This complex network of protein-protein interactions is however poorly characterized and the identification of its components should improve our understanding of APP biology and fate, and might enable the delineation of therapeutic approaches.

Given this context, we decided to focus on ADAMs and related proteins. Our study was inspired by several observations besides our knowledge of the involvement of ADAMs as α-secretases (Rosenberg, 2009): (i) ADAMs and APP are involved in many different biological processes including brain development, plasticity and repair (Yang et al., 2006), and (ii) several matrix metalloproteases (MMP-2, -3 and -9) can degrade Aβ peptides (White et al., 2006; Reitz et al., 2010; Carson and Turner, 2002). We therefore performed a multi-angle screen for new components of APP metabolism, with a focus on MMPs, ADAMs and related proteins. We postulated that differentially expressed MMPs, ADAMs and related proteins (when comparing expression in Alzheimer's disease brains and control brains) might be clues for their involvement in APP physiology.

## 2. Materials and Methods

Written informed consent was obtained from study participants or, for those with substantial cognitive impairment, from a caregiver, legal guardian, or other proxy and the study protocols for all populations were reviewed and approved by the appropriate Institutional review boards of each country.

All animal experiments were approved by the local animal care and use committee (*Comité d'Ethique en Experimentation Animale du Nord - Pas de Calais*, Lille, France).

### 2.1. Study Design

We postulated that uncharacterized MMPS, ADAMs or related proteins may be involved in the APP metabolism. The purpose of this study was thus to explore this possibility. Potential candidates were selected from transcriptomic analyses targeting MMPs/ADAMs expression using total RNAs extracted from the brain of AD cases and controls. The strongest variations in expression were validated in an independent sample of brains using a different technology. Potential correlation between amyloid deposition in the brain of AD cases and expression of our genes of interest were examined. This work allowed us to select ADAM30 for further exploration.

We developed ADAM30 over- or under-expression experiments in different cellular models to assess ADAM30 impact on the APP metabolism. Potential α-, β- or γ-secretase activities of ADAM30 were examined. A without a priori research for ADAM30 substrates was performed using COFRADIC experiments. Impact of ADAM30 on APP metabolism through CTSD activation was tested using pharmacological or siRNA tools. All the experiments have been made at least in triplicates and by two independent manipulators for most of them. This work allowed us to demonstrate that ADAM30 modulates the APP metabolism through CTSD activity.

The in vitro observations were finally extended to an "Alzheimer-like" transgenic mouse model specifically over-expressing ADAM30 in neurons. Primary cultures of adult neurons were used to validate the results obtained from cell lines. Measurement of soluble Aβ42 and amyloid deposition were performed to corroborate the in vitro results in the mouse brains. Electrophysiological analyses were finally performed to extend the results to neuronal activity. All the analyses were performed in a blinded fashion. Our data demonstrated that ADAM30 over-expression led to a decrease in Aβ42 secretion in primary cultures, in soluble Aβ42 and amyloid deposition in the cerebral tissue and to a rescue of LTP in the Alzheimer-like mouse brain.

### 2.2. Microarray Analyses

The preparation of human brain samples is described in the Supplementary experimental procedures. Total RNA was extracted with a phenol/chloroform protocol (TRIzol® reagent, Invitrogen®, USA) from frozen frontal cortex brain tissue from one hundred fourteen Alzheimer's disease samples and one hundred sixty seven control samples. The quality of the total RNA extract was assessed with an Agilent 2100 Bioanalyzer (Agilent) and the ribosomal 28S/18S RNA ratio was estimated with the system's onboard biosizing software. Twelve Alzheimer's disease cases and twelve controls were selected among the initial samples by applying the following two criteria: (i) a ribosomal 28S/18S RNA ratio of 1.0 or more; (ii) a Braak stage below 2 (for control samples) (Table S1) (Bensemain et al., 2009).

Specific oligonucleotides for one hundred thirty two open reading frames (corresponding to MMPs, ADAMs, ADAMTSs and related proteins) were designed using OLIGOMER software (Mediagen) (Supplementary experimental procedures and Table S2; Bensemain et al., 2009). After synthesis, the oligonucleotides were purified to obtain a population that was homogeneous in terms of length (Sigma-Aldrich®, Germany). All oligonucleotides were 5′-functionalized with a $C_6H_{12}NH_2$ arm.

In order to decrease the potential influence of inter-individual variability in the control population, we compared the genetic expression of each Alzheimer's disease case with the pool of control samples (see the Supplementary experimental procedures). Complementary RNA corresponding to 10 μg of the initial mRNA extract was produced by amplification and labeled with a Cy5 or Cy3 fluorophore using a Fluorescent Linear Amplification Kit® (Agilent), according to the manufacturer's instructions. After hybridization, microarrays were analyzed as described previously (Bensemain et al., 2009, see the Supplementary experimental procedures). It is important to note that the analyses of the one hundred thirty two ADAMs, MMPs and related proteins were performed with a one thousand seven hundred seventy one gene microarray used in a previous project in our laboratory (Bensemain et al., 2009; Chapuis et al., 2009). Accordingly, the threshold for statistical significance was set to $p < 10^{-5}$ (after Bonferroni correction), in order to select genes differentially expressed in the Alzheimer's disease brain.

### 2.3. Quantification of ADAM17, ADAM30, ADAM33 and ADAMTS16 mRNAs

Total RNA samples (from forty two controls and fifty one Alzheimer's disease cases) were randomly selected (see Supplementary material section) from among the samples not used in the transcriptomic experiments. Levels of ADAM17, ADAM30, ADAM33 and ADAMTS16 mRNAs were quantified as described by the supplier (Quantigene®, Panomics) (Canales et al., 2006), normalized against two different housekeeping genes (β-actin and β-glucuronidase: see the Supplementary material procedures) and were expressed in arbitrary units (AU). Comparisons were performed with a Mann-Whitney non-parametric test.

### 2.4. Immunohistochemistry Experiments

The brains used for immunohistochemistry (IHC) experiments were obtained at autopsy (at Lille University Medical Center) from Caucasian patients suffering from Alzheimer's disease and from Caucasian controls in whom the absence of Alzheimer's disease had been confirmed neuropathologically. These samples were not used in the transcriptomic experiments. In all patient samples, the neuropathologic features of Alzheimer's disease were confirmed by IHC and Western blot analysis of Tau, Aβ and α-synuclein (Delacourte et al., 2002). Brain samples were fixed in formalin for brightfield microscopy. Paraffin sections from the anterior frontal cortex (Brodmann area 10) were processed in an automatic Benchmark-XT system (Ventana, USA). Polyclonal antibodies against a sixteen-amino acid polypeptide within the human ADAM30 protein (Table S3) were raised according to a standard protocol (with three months of immunization, Interchim®, France). Anti-ADAM30 antibody and pre-immune rabbit serum (both diluted at 1:500) were applied after sample heating and were revealed using a standard immunoperoxidase technique.

### 2.5. Analyses of APP Metabolism as a Function of $ADAM30^{WT}$ or $ADAM30^{mut}$ Expression

Plasmid constructions are fully described in the Supplementary experimental procedures.

SKNSH-SY5Y-$APP^{695WT}$ and HEK293/HEK293-$APP^{695WT}$ cell lines were respectively cultured in DMEM and DEM/F12 supplemented with 10% serum FCS, 2 mM L-glutamine, 50 U/ml penicillin, 50 μg/ml streptomycin and (for SKNSH-SY5Y-$APP^{695WT}$ only) 1% MEM NEAA (Invitrogen®, USA) at 37 °C in a humidified atmosphere with 5% $CO_2$.

For the analysis of APP metabolism, SKNSH-SY5Y-$APP^{695WT}$ and HEK-$APP^{695WT}$ cell lines (300,000 cells per well at seeding) were transfected with pcDNA-Mock, pcDNA-$ADAM30^{WT}$, pcDNA-$ADAM30^{mut}$, shRNA pU6/ADAM30 vectors (500 μg) or siADAM30 (ON-TARGETplus ADAM30 siRNA, GE Dharmacon). Transient transfection was performed using an Exgen500 protocol (Fermentas®, France) in SKNSH-SY5Y-$APP^{695WT}$ cells and a Fugene-HD protocol (Roche Diagnostics®, Switzerland) in HEK/HEK-$APP^{695WT}$ cells, according to the respective manufacturers' instructions. SiADAM30 were specifically transfected using Lipofectamine RNAi Max. Of note, because of the low endogenous level of ADAM30 expression, shRNA and siRNA targeting ADAM30 were first validated in SKNSH-SY5Y stably over-expressing ADAM30 (Fig. S12).

Forty-eight hours after transfection, supernatants were collected for the quantification of APP by-products and cells were lysed to assay for expression of $ADAM30^{WT/mut}$ or endogenous ADAM30 by Western blotting. Aβ1–40, Aβ1–42, sAPPα and sAPPβ were measured in sandwich ELISAs (INNOTEST® β-amyloid (1–42) and β-amyloid (1–40), Innogenetics, Ghent, Belgium), and the sAPPα and sAPPβ Assay kits (IBL-International®, Germany), according to the respective manufacturers' recommendations. Three independent, duplicate experiments were carried out for each mutant and measurements were performed twice for each sample.

For the BACE1 fluorometric assay, HEK293 cells were transfected with $ADAM30^{WT/mut}$ cDNAs. Forty-eight hours after transfection, cells were pipette-lysed with 50 μl of Tris-HCl (10 mM, pH 7.5) and monitored for BACE1 activity, as described previously (Andrau et al., 2003). For the γ-secretase activity assay, solubilized membranes were obtained from intact HEK293 cells transfected with $ADAM30^{WT/mut}$ cDNAs and analyzed as described previously (Sevalle et al., 2009).

For analyses of $APP\Delta C8$, $APP^{F690S}$ and $APP^{E691V}$ metabolism, HEK293 cell lines (300,000 cells per well at seeding) were transfected with pcDNA-Mock or pcDNA-$ADAM30^{WT}$ (250 μg) and pcDNA-$APP^{695WT}$, pcDNA-$APP^{\Delta C8}$, pcDNA-$APP^{F690S}$ or pcDNA-$APP^{E691V}$ (250 μg). Three independent, duplicate experiments were carried out for each mutant and all measurements were performed twice for each sample.

### 2.6. Identification of ADAM30 Substrates

Two HEK293-$APP^{695WT}$ cell lines stably overexpressing $ADAM30^{WT}$ or $ADAM30^{mut}$ (HEK293-$ADAM30^{WT}$ and HEK293-$ADAM30^{mut}$) were generated (as mentioned above, one advantage of the HEK293 cell line is the absence of endogenous ADAM30 expression; see the Supplementary experimental procedures). These HEK293 cell lines were used to perform a targeted COFRADIC analysis of N-terminal peptides in a highly complex mixture, whereas all internal peptides were disregarded. The combination of COFRADIC, SILAC and N-terminal tagging chemistries generates quantitative data on the modification status of the N termini of the proteins present in the mixture. This approach provides an overall profile of enzyme activities and substrates (Staes et al., 2011; see the Supplementary experimental procedures).

For the CTSD assay, recombinant $ADAM30^{WT}$ and $ADAM30^{mut}$ were generated by Genscript (Piscataway, USA) (see the Supplementary experimental procedures). The activity of recombinant human CTSD (R&D Systems, Lille, France) was measured via cleavage of a fluorogenic peptide substrate (Mca-PLGL-Dpa-AR-$NH_2$ from R&D Systems). CTSD (16 μg/ml) was pre-incubated at 37 °C for 30 min before being diluted eight-fold into 50 mM Tris-HCl, pH 7.5 buffer with 300 mM NaCl. Twenty-five microliter aliquots of enzyme sample were transferred into the wells of black 96-well microplates (Corning, Amsterdam, The Netherlands) and incubated with substrate at 37 °C for 40 min. The final volume of the incubation solution was 100 μl and the final reagent concentrations were 0.5 μg/ml of CTSD and 10 μM of substrate in assay buffer (0.1 M sodium acetate and 0.2 M NaCl, pH 3.5). Fluorescence was measured using a Victor3 V1420 plate reader (Perkin Elmer,

Courtaboeuf, France) fitted with a 340 nm excitation filter and a 410 nm emission filter. The fluorescence of substrate blanks was subtracted. To evaluate the effect of ADAM30$^{WT}$ or ADAM30$^{mut}$ on CTSD activity, CTSD was pre-incubated with 84 μg/ml of each protein prior to dilution and substrate addition. The observed CTSD activities were compared with that measured when CTSD was pre-incubated with 84 μg/ml of bovine serum albumin (Sigma, Lyon, France).

For measuring intracellular CTSD activity, the SKNSH-SY5Y-APP$^{695WT}$ and HEK-APP$^{695WT}$ cell lines (300,000 cells per well at seeding) were transfected with pcDNA-Mock, pcDNA-ADAM30$^{WT}$ or pcDNA-ADAM30$^{mut}$ vectors. After 48 h, cells were lysed and the time course of CTSD activity (0, 5, 10 and 15 min) was monitored using a fluorometric SensoLyte® Cathepsin D assay (AnaSpec®, USA).

For the cathepsin inhibition experiments, HEK-APP$^{695WT}$ cell lines were transfected for 24 h with pcDNA-Mock, pcDNA-ADAM30$^{WT}$ or pcDNA-ADAM30$^{mut}$ vectors and then exposed for 24 h to pepstatin A (10 μM; Sigma-Aldrich, Germany), leupeptin (10 μM; Sigma-Aldrich, Germany) or E64-D (1 μM; Sigma-Aldrich).

Transient Co-transfection of CTSD siRNA (37.5 pmol ~0.75 μg, ON-TARGETplus CTSD siRNA, GE Dharmacon) and pCDNA3.1-ADAM30 vector (1.25 μg of DNA per well) was performed using Lipofectamine® 2000 Transfection Reagent (ThermoFisher Scientific) according to the manufacturer's recommendations. After 36 h of transfection and 24 h of secretion (wash after 12 h of transfection), cells were finally lysed and supernatants were recovered. Cell extracts were analyzed by western blot to validate both ADAM30$^{WT}$ over-expression and CSTD under-expression.

By-products of APP metabolism were then quantified as previously described. Three independent duplicate/triplicate experiments were carried out and measurements were performed twice for each sample.

### 2.7. Western Blot

Antibody against CTSD was diluted at 1/5000. The primary antibodies used to detect ADAM30 were diluted at 1/1000. The antibody 6E10 was used for APP detection at 1/5000. A monoclonal mouse antibody was used to detect β-actin at 1/10000. Antibody references are indicated in Table S3. Immunoreactive complexes were revealed using the ECL™ Western Blotting kit (Amersham®). Membranes were digitized using the ChemiDoc MP System (Bio-Rad, Marnes-la-Coquette, France).

### 2.8. Immunofluorescence and PLAs

The SKNSH-SY5Y-APP$^{695WT}$ cell line was cultured on poly-L-Lys-coated glass coverslips (Lab-Tek® Chamber Slide System 2 wells, Nunc, Roskilde, Denmark) for 24 h. The cells were then transfected with pcDNA-ADAM30$^{WT}$-GFP, pcDNA-ADAM30$^{mut}$-GFP, pcDNA-ADAM30$^{WT}$ or pcDNA-ADAM30$^{mut}$ vectors. After 48 h, cells were fixed in PBS containing 4% paraformaldehyde for 20 min at room temperature and further permeabilized with 0.25% (v/v) Triton X-100 in PBS. After blocking in 5% (w/v) BSA, fixed materials were incubated overnight at 4 °C with primary antibodies (diluted 1/500 in PBS supplemented with 5% (w/v) BSA and 0.25% Triton X-100.). After washing, appropriate secondary antibodies were used (diluted 1/400). Primary and secondary antibody combinations are described in Supplementary Table 3. The slides were read under a confocal microscope (Leica LSM 710) at the MICPAL microscopy facility at the Pasteur Institute of Lille (Lille, France).

For the PLAs, cells were washed with PBS and then fixed in PBS containing 4% paraformaldehyde for 30 min at room temperature. Cells were permeabilized with 0.25% Triton X-100 in PBS for 10 min. After blocking in 1% bovine serum albumin (BSA), cells were incubated for 2 h at room temperature with primary antibodies diluted in 1% BSA in PBS. The *anti*-CTSD antibody (Abcam, catalog number ab6313-100) was diluted 1/100 and the anti-ADAM30 antibody (Genetex, catalog number GTX117694) was diluted 1/50. Cells were then washed three times in PBS. All reagents used in the PLAs were purchased from Olink Bioscience (Uppsala, Sweden). The PLAs were performed according to the manufacturer's instructions by using anti-CTSD and anti-ADAM30 as primary antibodies.

### 2.9. Transgenic Mouse Experiments

#### 2.9.1. Mouse Models

Two conditional ADAM30$^{floxstopflox}$ and ADAM30$^{mutfloxstopflox}$ transgenic mice were generated within a C57Bl6N background (Taconic, Germany). In brief, a construct containing the GAGGS promoter, a *LoxP*-NeomycineStop-*LoxP* cassette and the human Adam30$^{WT}$ or Adam30$^{mut}$ gene was introduced In the Rosa26 locus. CamKIIα/Cre mice (in which Cre gene expression is driven by the CamKIIα promoter (Tsien et al., 1996) and hAPP$^{swe,Ind}$ mice expressing a human APP gene bearing Swedish (670/671KM-NL) and Indiana (717 V-F) mutations were obtained from The Jackson Laboratory (Mucke et al., 2000). Both mice have a C57Bl6J background. The first cross was between hAPP$^{swe,Ind}$ mice and CamKIIα-Cre mice, yielding APP$^{+/-}$/Cre$^{+/-}$ mice. The second cross was between APP$^{+/-}$Cre$^{+/-}$ mice and hADAM30$^{floxstopflox+/-}$ or hADAM30$^{mutfloxstopflox+/-}$ mice, yielding triple transgenic hADAM30-$^{WTΔstop}$-hAPP$^{swe,Ind}$-Cre or hADAM30$^{mutΔstop}$-hAPP-Cre mice expressing ADAM30 or ADAM30$^{mut}$ in their brains. The mice data were systematically obtained from a crossing of a hADAM30$^{floxstopflox+/-}$ heterozygous C57BL6N female with an APP$^{+/-}$ Cre$^{+/-}$ heterozygous C57BL6J male without any backcross. So the genetic background has been systematically controlled and the phenotypes were analyzed only at the first generation.

All the mice were genotyped by PCR using genomic DNA isolated from tail tips, according to the provider's protocols (The Jackson Laboratory; Table S4 and Fig. S13), Mice were maintained on a standard diet with ad libitum access to water in a specific pathogen-free animal facility. The mice were studied at the age of 10 months ± 1 week. At sacrifice, the brains were removed. After sagittal section, one hemibrain was used for histological studies and the other was sectioned on brain matrices to isolate the cortices and hippocampi, which were then stored at −80 °C.

#### 2.9.2. Aβ42 Peptide Assays

Following the validation of gene expression in the brain (see the Supplementary experimental procedures), snap-frozen hippocampi were homogenized in guanidine buffer (Johnson-Wood et al., 1997). Hippocampal levels of Aβ42 were quantified with commercially available ELISA kits (Innogenetics kit for Aβ42). The protein concentration in hippocampal extracts was determined using a Bradford assay kit (Bio-Rad Laboratories).

#### 2.9.3. Measurements of CTSD Activity Ex Vivo

Cortex samples were homogenized with dithiothreitol-containing assay buffer from the SensoLyte 520 Cathepsin D assay kit (AnaSpec®, USA), using a high-power homogenizer. After centrifugation for 10 min at 12,000 rpm and 4 °C, supernatants were assayed for CTSD with the SensoLyte® Cathepsin D assay kit (AnaSpec®), according to the manufacturer's instructions. The CTSD activity was measured independently in two different experiments: a time-course measurement every 5 min between 0 min and 30 min and an end-point measurement at 30 min (Fig. 4). Each measurement was performed at least twice for each cortex sample and the results were normalized against the sample weight and protein concentration. At least three independent, duplicate experiments were carried out and measurements were performed twice for each sample. Comparisons were performed using a Mann-Whitney non-parametric test (for the end-point experiment) or a two-tailed *t*-test (for the time-course experiment).

#### 2.9.4. Histology

Hemibrains from APP$^{sw,Ind}$, ADAM30$^{WT}$/APP$^{sw,Ind}$ and ADAM30$^{WT}$/APP$^{sw,Ind}$/Cre mice were processed by a high-throughput neurohistological service (NeuroScience Associates). Briefly, hemibrains were embedded in a green-colored solid matrix (Multibrain technology) and cut along the rostrocaudal axis. For each brain, three sets of 30-μm-thick coronal sections (one hundred per set) were collected. One set was used for Aβ-peptide IHC experiments with a 6E10 monoclonal antibody and 3,3′-diaminobenzidine development. The other two series were stored for further analysis. Block-face photographs were taken (lateral resolution: 13 μm) before each section was prepared (Canon EOS 5D Mark III) (Dubois et al., 2007). Histology images were digitized using a flatbed scanner (ImageScanner III, GE Healthcare) with a lateral resolution of 5 μm.

#### 2.9.5. 3D Reconstruction

All the image processing steps described below were performed using the BrainRAT pipeline 3D-HAPi (Dubois et al., 2007, Dubois et al., 2010; Vandenberghe et al., 2016) and a dedicated in-house image processing software package (BrainVISA, http://brainvisa.info). All image analyses were performed by operators blinded to the animal group assignments (Fig. S14). For each brain, block-face photographs and histological images were stacked. Brain tissue was automatically segmented on block-face photography volumes and masked to remove background. As block-face images were taken prior to sectioning at the same position section after section, reconstructed photographic volumes necessarily reflected the original shape of the frozen brains and were used as a 3D reference. Affine registrations between two-dimensional histological images and corresponding block-face photographs were estimated and used to (i) correct for deformations due to histological procedures and (ii) provide coherent histological volumes (xyz resolution: $5 \times 5 \times 90$ μm$^3$ for histological volumes stained with the 6E10 monoclonal antibody). For each animal, the block-face photography volume and the corresponding IHC volume stained with 6E10 monoclonal antibody were aligned within the same spatial reference framework.

#### 2.9.6. Quantification of the Amyloid Load

Aβ peptide aggregates were segmented from the rest of the image by applying a supervised Bayesian classifier (Chubb et al., 2006) to reconstructed 6E10 IHC volumes. Each voxel was classified into either positively stained amyloid plaques, non-stained tissue or background as a function of the color intensities (red, green and blue) and the mean neighborhood intensity (the mean red, green and blue intensities within the voxel's four-connected neighbors). A mouse hemibrain atlas was derived from a publicly available, magnetic resonance imaging (MRI)-based mouse brain atlas (http://www.mouseimaging.ca/technologies/C57Bl6j_mouse_atlas.html) (Dorr et al., 2008) and registered on each photographic volume according to a protocol described and validated elsewhere (Lebenberg et al., 2010). Briefly, non-linear transformations were estimated for each brain (using MRI) and then applied to corresponding label atlas volumes. The percent volume occupied by the 6E10 staining was computed for the forebrain as a whole and the hippocampus in particular. Quality controls were performed visually for both registration and color segmentation processes. Comparisons between amyloid loads were performed with a Mann-Whitney non-parametric test.

#### 2.9.7. Electrophysiology Experiments

Mice were shipped to E-PHY-SCIENCE (Sophia-Antipolis, France). In this study, twenty six 12-month-old transgenic mice (hADAM30$^{WT}$-hAPP$^{Sw,Ind}$-Cre, hADAM30$^{WT}$-hAPP$^{Sw,Ind}$, hAPP$^{Sw,Ind}$) and 8 WT mice were used. In total thirty four mice were scarified for the study. Of note, premature death is a phenotype of transgenic Alzheimer mice and no mice were lost in any of the groups during the course of the study. Acute slices (400 μm thick) were prepared with a vibratome (VT 1000S; Leica Microsystems, Bannockburn, IL) in ice-cold gassed aCSF enriched with sucrose. Sections were incubated in aCSF supplemented at 34 °C for 20 min and then kept at room temperature for at least 1 h before recording.

Recording was performed in a submerged chamber continuously perfused with aCSF, at 1–2 ml/min. A monopolar electrode was placed in the Schaffer collaterals, and stimulation was applied at 0.066 Hz (every 20 s) with stimulus intensity ranging from 0 to 100 μA, yielding evoked field EPSPs (fEPSPs) of 0–0.3 V.

fEPSPs were recorded in the stratum radiatum using a borosilicate micropipette filled with aCSF.

The signal was amplified with an Axopatch 200B amplifier (Molecular Devices, Union City, CA), digitized by a Digidata 1200 interface (Molecular Devices) and sampled at 10 kHz with Clampex 10 (Molecular Devices). aCSF, during incubation and recording, was composed of the following (in mM): 119 NaCl, 11 D-glucose, 1.3 MgCl2·6H2O, 1.3 NaH2PO4, 2.5 KCl, 2.5 CaCl2, 26 NaHCO3, gassed with O2/CO2 (95/5%) at least 20 min before use and throughout the experiment. Baseline was recorded for a minimum of 20 min or until stable. LTP was induced by stimulation with 100 Hz with four trains of a 1-s tetanus separated by 20 s.

Recordings were acquired using Clampex (Molecular Devices) and analyzed with Clampfit (Molecular Devices). Experimenters were blinded to treatment for all experiments. One or more slices from each mouse were used and were averaged, so that animals and not slices are considered biological replicates. Data were analyzed by measuring the slope of individual fEPSPs at 1–1.5 ms after the stimulus pulse by linear fitting using Clampfit (Molecular Devices).

### 3. Statistical Analyses

Statistical analysis was performed using SAS statistical software (version 9.1, SAS Institute Inc., Cary, NC, USA).

### 4. Results

#### 4.1. Transcriptomic Analyses Show That ADAM30 is Under-expressed in AD Brains

Using total RNA from the frontal cortex of twelve Alzheimer's disease cases and twelve healthy controls, we performed a transcriptomic analysis of one hundred thirty two genes coding for MMPs, ADAMs and related proteins (the complete results are provided in Table S2). Four ADAMs were found to be differentially expressed in the brain of Alzheimer's disease cases compared with controls (ADAM17, ADAM33, ADAMTS16 over-expressed and ADAM30 underexpressed; $p < 1 \times 10^{-5}$ for all comparisons after Bonferroni correction; see the Materials and Methods section and in Bensemain et al. (2009)). Of note, since the analyses of the 132 ADAMs, MMPs and related proteins were performed using a 2741-gene homemade microarray (see again the Materials and Methods section and in Bensemain et al. (2009), this design allowed us to determine whether the decrease in ADAM30 expression may be imputed to neuronal death. In order to evaluate this potential bias, we compared the expression of a subset of genes reported as only expressed in neuron (n = 113) to genes not exclusively (or not) expressed in neurons (n = 1789). Importantly, we did not observe any significant differences in terms of fold-change (Fig. S1). To bypass inherent risks of false positives in such systematic transcriptomic approaches despite of multiple testing corrections, the expression level of these four ADAMs was then analyzed in an independent, larger sample of AD cases (fifty one) and healthy controls (forty two). ADAM30 underexpression (Fig. 1a) and ADAM33 overexpression were confirmed whereas changes in the expression of ADAM17 and ADAMTS16 genes could not be validated in this larger sample (Fig. S2).

We finally assessed whether the levels of ADAM30/ADAM33 expression may be correlated with AD hallmarks in the brain. The decrease in ADAM30 expression was significantly correlated with higher Aβ42 loads but not with Tau loads in Alzheimer's disease brain samples (Fig. 1b and Fig. S3) while ADAM33 expression did not correlate with these Alzheimer's disease markers (data not shown). Immunohistochemistry experiments in human brain tissue revealed a neuronal expression of ADAM30 (Fig. 1c and Fig. S4) as observed in a laser dissection transcriptomic analysis (GSE15222 dataset described in Liang et al. (2008)). We thus hypothesized that ADAM30 under-expression might be harmful by modulating Aβ peptide production and thus, we assessed the putative involvement of ADAM30 in APP processing.

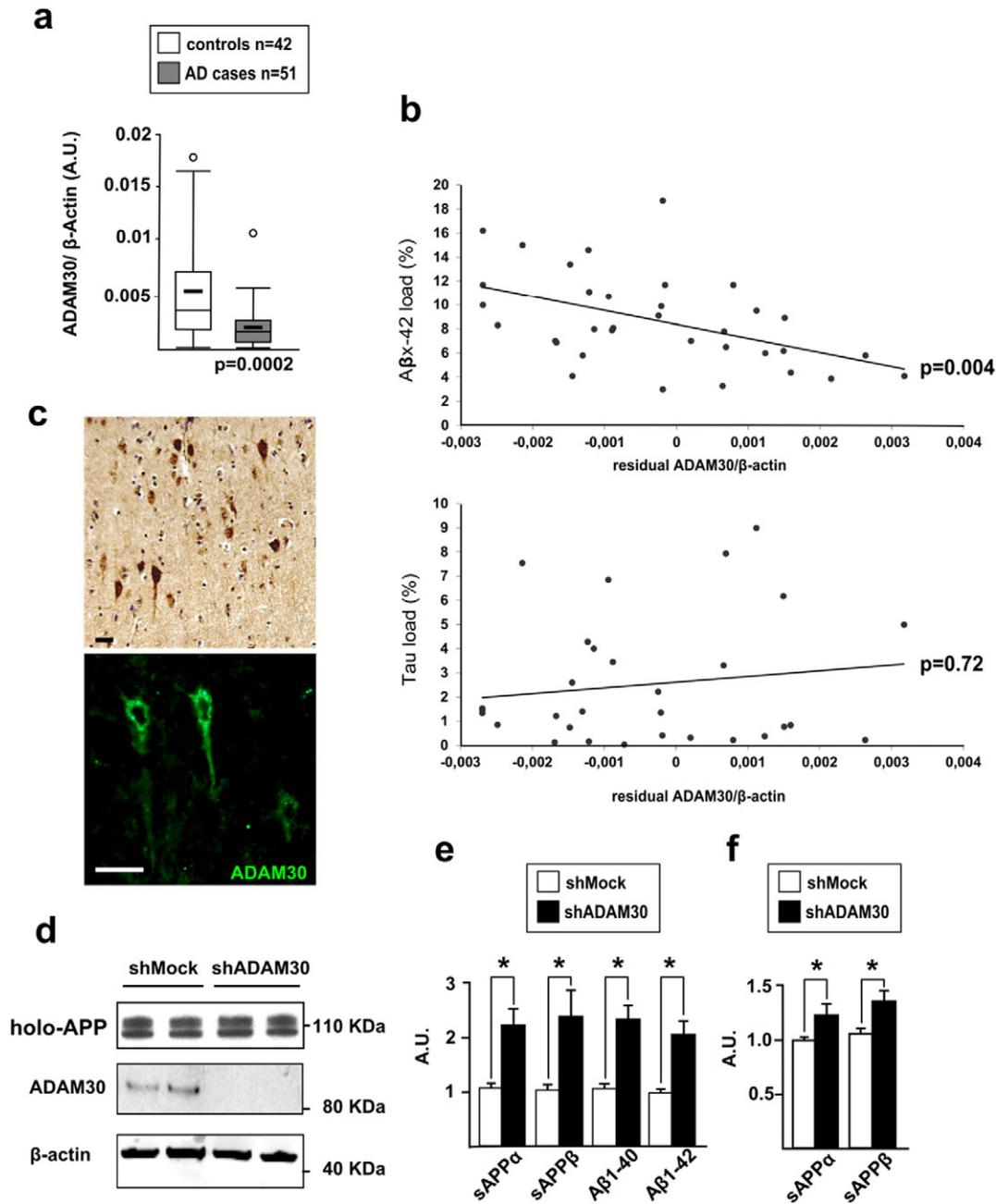

**Fig. 1.** Characterization of ADAM30 expression in the brain. (a) Expression levels of *ADAM30* in the brains of Alzheimer's disease cases (fifty one) and controls (forty two). All values are reported as arbitrary units (AU) following normalization against β-actin mRNA levels. All quantifications were carried out in triplicate in all individuals. The thick lines represent the median *ADAM30* expression level in cases and controls. The midline represents the mean value and the upper and lower horizontal lines represent the first and third quartiles, respectively. Circles indicate individuals with extreme values (more than 2 SD above or below the mean value). p-Values refer to a Mann-Whitney non-parametric test. (b) Association of Aβx-42 and Tau loads in the brain of Alzheimer's disease cases with the expression of ADAM30 (residual correction), normalized against the expression of a-actin housekeeping gene. p values refer to Spearman's non-parametric test. (c) Immunohistochemistry experiments in human brain supporting ADAM30 expression in neurons. (d) A representative experiment measuring transfection of a shRNA vector against ADAM30 into SKNSH-5Y5Y-APP695$^{WT}$ cells (ADAM30, and β-actin). (e) Mean differences (±SEM) in the amounts of sAPPα, sAPPβ, A-1β40, A-1β42 in SKNSH-Sy5Y-APP695$^{WT}$ cells or (f) endogenous sAPPα, sAPPβ in SKNSH-SY5Y. Three independent experiments were performed in duplicate in SKNSH-5Y5Y-APP695$^{WT}$ and in triplicate in SKNSH-SY5Y. *p < 0.05 (Mann-Whitney non-parametric test).

### 4.2. ADAM30 Under-expression is Associated With Increased APP Catabolites In Vitro

We first investigated whether modulation of ADAM30 expression could be associated with an alteration of the APP metabolism in the SKNSH-SY5Y cell line stably expressing the wild-type (WT) APP$^{695}$ isoform (SKNSH-SY5Y-APP$^{695WT}$). This model allows measuring the production/secretion of all APP byproducts and quantifying separately Aβ$_{1-40}$ and Aβ$_{1-42}$ peptides. ADAM30 under-expression (transient transfection of SKNSH-SY5Y-APP[695WT] cells with a short hairpin RNA (shRNA) targeting ADAM30; Fig. 1d)) increased the levels of all APP products yielded by α-, β- and γ-secretases-mediated proteolysis and particularly Aβ$_{1-40}$ and Aβ$_{1-42}$ (Fig. 1e). Both endogenous sAPPα and sAPPβ secretion were also increased in the SKNSH-SY5Y cells after transient transfection of ADAM30-shRNA (Fig. 1f) or of an ADAM30-siRNAs (data not shown).

### 4.3. ADAM30 Catalytic Activity is Required for the Modulation of APP Metabolism

ADAM30 holds a unique zinc-binding motif HEXXHXXGXXHD, which is normally required for enzymatic activity (all metalloproteases harbour a HEXXH motif and half of the ADAM proteins present such a functional catalytic motif). We thus aimed at determining whether the catalytic function of ADAM30 accounts for the observed modulation of APP catabolites. To assess this possibility we generated mammalian expression vectors expressing either a wild-type ADAM30 (ADAM30[WT]) or an ADAM30 with a mutated catalytic site (ADAM30[mut], see Supplementary information). These constructs were transfected either in the SKNSH-SY5Y-APP[695WT] cell line or a HEK293 cell line also stably expressing the wild-type APP[695] isoform (HEK293-APP[695WT]).

Over-expression of ADAM30[WT] in both cell lines triggers the exact opposite effects observed when down-regulated ADAM30, i.e. decreased levels of all amyloidogenic and non-amyloidogenic APP catabolites (Fig. 2A and 2B). This effect was not observed after overexpression of catalytically silent ADAM30[mut] (Fig. 2a and b).

We next started evaluating pathways that could link ADAM30 and APP catabolites. Strikingly, the fact that all APP catabolites were decreased by ADAM30 could have indicated that ADAM30 acted upstream of secretases cleavages. One possibility could be that ADAM30 directly modulates APP levels but we showed that APP mRNA levels were not affected by ADAM30 modulation (Fig. S5). An alternative explanation would be that ADAM30[WT] (but not ADAM30[mut]) modifies α-, β- and γ-secretase activities. Since several ADAMs has been proposed as genuine α-secretases, we first examined this possibility. However, ADAM30 did not exhibit α-secretase-like activity (Fig. S6a). Of note, ADAM30 is not present at the membrane (Fig. S6b, c). We also ruled out ADAM30 as a modulator of β- and γ-secretases, since overexpression of WT or mutated ADAM30 did not modify these activities (Fig. S7).

### 4.4. Identification of ADAM30 Substrates and Subsequent Targets

The above data suggest that ADAM30 might modulate an intracellular pathway or an intermediate effector contributing to APP physiology. Thus, we systematically searched for direct ADAM30 substrates or further downstream targets by applying *N*-terminal combined fractional diagonal chromatography (COFRADIC) (Staes et al., 2011) and comparing the "stable isotope labeling with amino acids in cell culture" (SILAC)-labeled *N*-terminones of HEK293 cells stably over-expressing either WT or catalytically inactive ADAM30 (Fig. 2b). We identified two thousand two hundred thirty eight proteins, eighteen of which (characterized by twenty five distinct peptide sequences) had neo-*N*-terminal peptides with higher intensities in cells over-expressing ADAM30[WT], suggesting a putative susceptibility to ADAM30-mediated proteolysis (Fig. 2d). Most of these eighteen proteins were involved in transcription and translation (e.g. histones and ribosomal proteins). Even though the latter might well be substrates for ADAM30, we suspected that this observation might reflect potential differential extraction of proteins, e.g. potential differences in cell growth between HEK293-ADAM30[WT] and HEK293-ADAM30[mut] cell lines. After discarding all proteins linked to transcription and translation, only three peptides (and thus proteins) showed different ratios: insulin receptor substrate 4 (IRS4), cathepsin D (CTSD) and G kinase-anchoring protein 1 (GKAP1). We decided to focus on CTSD (see representative mass spectrum of the neo *N*-terminal peptide of CTSD, Fig. 2e). Indeed, the potential ADAM30 cleavage site identified in the COFRADIC experiment matched the CTSD endoproteolysis domain (residues 161–169) leading to a mature, active, lysosomal protease (Fig. 3A) (Benes et al., 2008). This latter observation suggests that ADAM30 acts as a CTSD maturation enzyme leading to enhanced cellular activity.

### 4.5. CTSD Co-localizes With, and is Activated by ADAM30

To further investigate the hypothesis whereby ADAM30 targets CTSD, we set up a fluorescence-based assay of CTSD activity. A human, recombinant pro-CTSD was incubated with its substrate in the presence or absence of human recombinant ADAM30[WT], ADAM30[mut] or bovine serum albumin (BSA). Incubation with ADAM30[WT] increased CTSD activity, whereas this effect was not detected after incubation with either ADAM30[mut] or BSA (Fig. 3b). In native SKNSH-SY5Y and transiently transfected HEK293-APP[695WT] cells, proximity ligand assay (PLA) revealed that respectively endogenous and overexpressed ADAM30[WT] co-localized with endogenous CTSD (Fig. 3c). Overexpression of ADAM30[WT] (but not ADAM30[mut]) was consistently associated with elevated CTSD activity in transiently transfected HEK293-APP[695WT] cells (Fig. 3d) or SKNSH-SY5Y-APP[695WT] cells (Fig. 3e). Inversely, transfection of ADAM30-siRNA led to a decrease in CTSD activity in native SKNSH-SY5Y cells (Fig. 3f). Of note, mutagenesis of ADAM30[mut] did not appear to alter the protein compartmentalization when compared with the main localization of ADAM30[WT] in late-endosome (Fig. S8).

### 4.6. CTSD Activity is Required for the Modulation of APP Metabolism by ADAM30

Both ADAM30's cellular localization and its ability to increase CTSD activity led us to postulate that ADAM30[WT] could contribute to the lysosomal CTSD maturation/activation and thereby, could favor APP degradation. This hypothesis may explain our observation of a general decrease in amyloidogenic and non amyloidogenic APP catabolites and agrees with our empirical observation that exposure to generic lysosome inhibitors (e.g. bafilomycin A1 and chloroquinone) abolished ADAM30[WT]-mediated impact on cellular APP metabolism (Fig. S9).

We further investigated this hypothesis by testing the impact of different pharmacological inhibitors: pepstatin (an aspartyl protease inhibitor with activity mainly directed towards CTSD), leupeptin (a serine and cysteine proteinase inhibitor with selective activity against cathepsin B) and E64D (a cysteine proteinase inhibitor that targets CTSB, cathepsin L and calpain). In the presence of pepstatin, the effect of ADAM30[WT] expression in HEK293 cells was no longer detected (Fig. 4a). In contrast, leupeptin (a serine and cysteine proteinase inhibitor with selective activity against CTSB) did not alter the impact of ADAM30[WT] expression on APP metabolism in HEK293-APP695[WT] cells (Fig. 4a). Similar results were obtained with E64D (Fig. 4a). Pepstatin could have elicited non-specific effect on the APP metabolism. In order to confirm the molecular cascade linking ADAM30, CTSD and APP, we finally down-regulated CTSD expression with a siRNA targeting CTSD in HEK293 cells over-expressing or not ADAM30[WT]. When CTSD was down-regulated, the effect of ADAM30$^{WT}$ expression in HEK293 cells was again no longer detected (Fig. 4b).

**Fig. 2.** Representative experiment measuring ADAM30[WT] and ADAM30[mut] overexpression in (a) SKNSH-5Y5Y-APP695[WT] and (b) HEK-APP695[WT] and mean differences (± SEM) in the secreted amounts of sAPPα, sAPPβ, A1-β40 and A1-β42. Three independent experiments were performed in duplicate. *p < 0.05 (Mann-Whitney non-parametric test). (c) HEK293 cell lines stably over-expressing ADAM30[WT] or ADAM30[mut] and cultured them in SILAC medium supplemented with 12C- or 13C-labeled arginine, respectively. After lysis, equal amounts of proteins from the two samples were mixed, conditioned and then analyzed in a two-step HPLC process (see the Materials and Methods section). (c) Eighteen proteins (characterized by twenty five distinct peptide sequences in COFRADIC) were significantly over-represented in the HEK293-ADAM30[WT] cell line, relative to the HEK293-ADAM30[mut] cell line. (d) A representative mass spectrum of the neo N-terminal peptide of CTSD (ALGGVKVER). The peptide with mass *m/z* 509,81 Da originates from the 12C6 arginine-labeled cell line (mutant cell line), whereas the peptide with mass *m/z* 512,82 Da originates from the 13C6 arginine-labeled cell line (WT cell line).

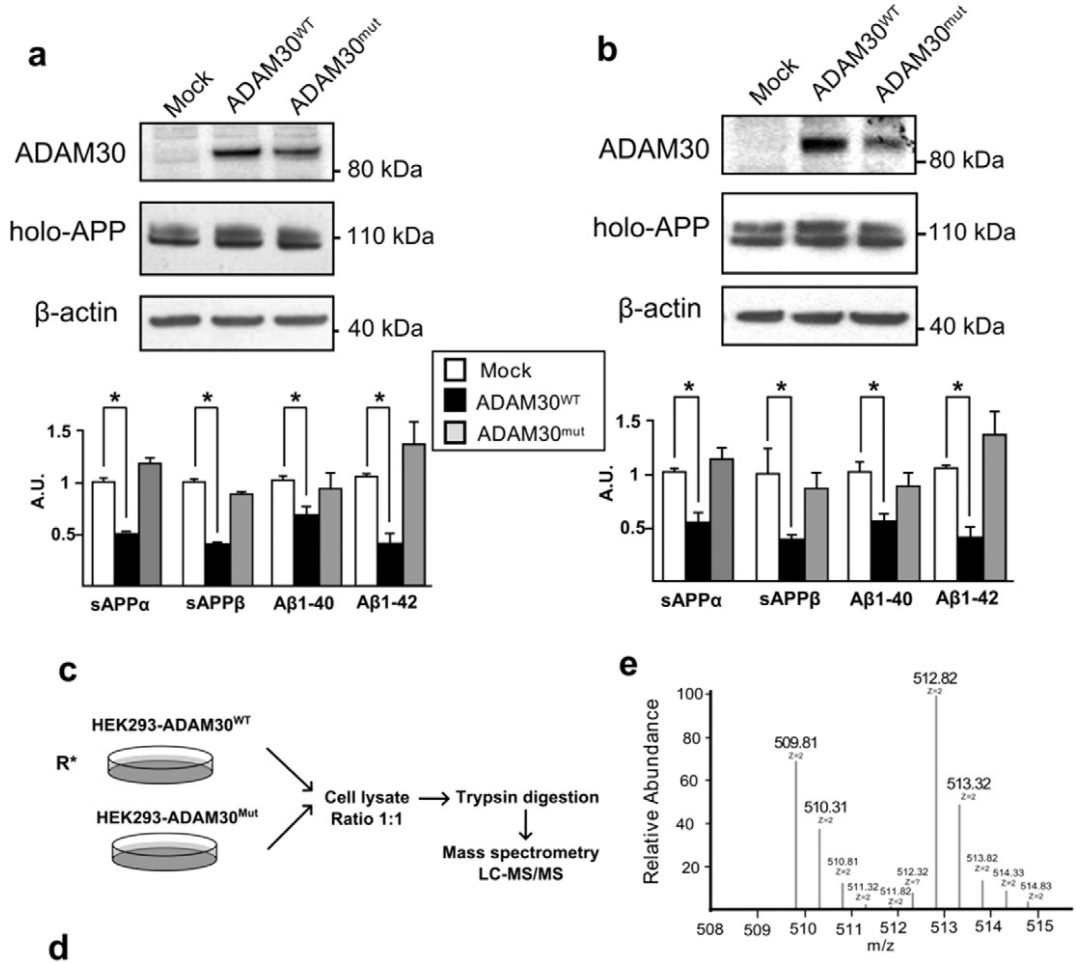

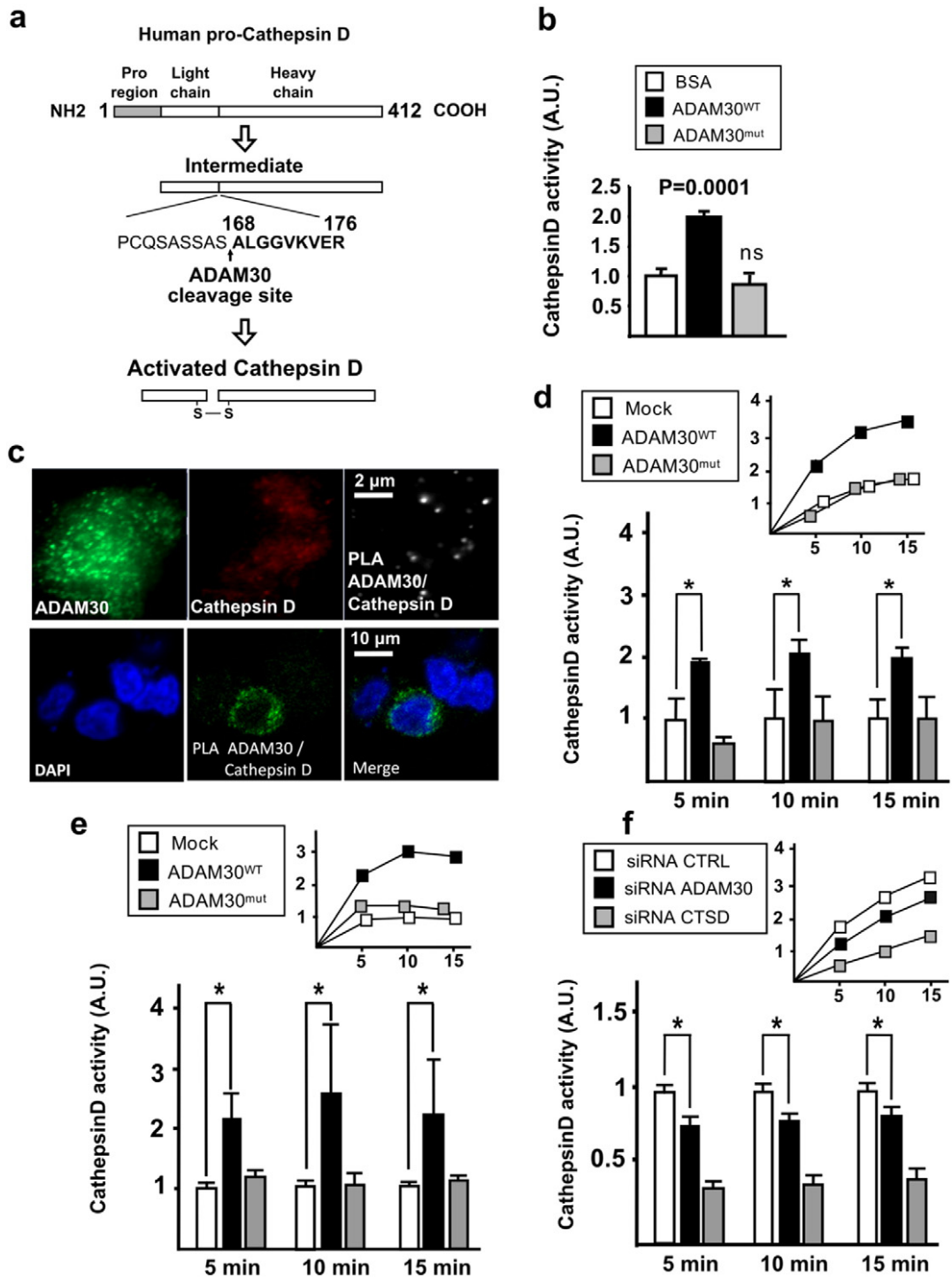

**Fig. 3.** Characterization of CTSD activation by ADAM30. (a) Proteolysis of the pro-CTSD for CTSD activation. The black arrow shows the cleavage site for ADAM30 between amino acids 167 and 168 (corresponding to the activational processing of CTSD). (b) In vitro measurement of CTSD activity (mean ± SD) after incubation of recombinant BSA, ADAM30$^{WT}$ or ADAM30$^{mut}$ with a CTSD peptide substrate. (c) Proximity ligation assay with human endogenous ADAM30$^{WT}$ and CTSD proteins in native SKNSH-SY5Y and with overexpressed ADAM30$^{WT}$ and endogenous CTSD in HEK293 cells. CTSD activity in HEK293-APP695$^{WT}$ (d) or SKNSH-SY5Y-APP695$^{WT}$ (e) cells transfected with ADAM30$^{WT}$, ADAM30$^{mut}$ or empty vector (Mock). The CTSD activity was assayed in three independent experiments, with a measurement every at 5, 10 and 15 min (means ± SEM). Top-right: representative kinetic experiment. (f) CTSD activity in SKNSH-SY5Y cells transfected with ADAM30-siRNA. The CTSD activity was assayed in three independent experiments, with a measurement every at 5, 10 and 15 min (means ± SEM). Top-right: representative kinetic experiment. *$p < 0.05$ (Mann-Whitney non-parametric test). Three independent experiments were performed in duplicate.

Altogether, these results strongly suggested that ADAM30 might be involved in APP degradation via selective activation of lysosomal CTSD. In line with our previous results, it is noteworthy that in a HEK293 cell line stably overexpressing APP$^{695WT}$ and ADAM30$^{WT}$, exposure to pepstatin was also associated with a 51% increase in intracellular APP accumulation (relative to cells not overexpressing ADAM30$^{WT}$; $p < 0.01$; Fig. 4c).

### 4.7. Lysosome Sorting of ADAM30 is Required for its Modulation of APP

We finally assessed whether the consensus sequences for lysosomal sorting in the APP C-terminal tail (Lai et al., 1995; Kouchi et al., 1998;

Kouchi et al., 1999) was required for ADAM30$^{WT}$ impact on APP metabolism. In a first step, we mutated lysine 688 of APP$^{695WT}$ into a stop codon (APPΔC8) (Fig. 4d); this leads to deletion of the last eight C-terminal amino acids (the lysosome-addressing sequence YKFF, corresponding to YXXØ, where Ø is a highly hydrophobic residue) but does not modify the sequence controlling APP endocytosis. Fig. 4e shows that the transient co-expression of APPΔC8 and ADAM30$^{WT}$ in HEK293 cells abolished the decrease in APP catabolites associated with ADAM30$^{WT}$ expression. Further, we mutated phenylalanine at position 690 (APP$^{F690S}$) into a serine residue within the lysosome-addressing sequence. Fig. 4e also shows that transient co-expression of APP$^{F690S}$ and ADAM30$^{WT}$ in HEK293 cells abolished the decrease in APP products associated with ADAM30[WT] expression. However, mutation of glutamic acid 691 (outside the lysosome-addressing sequence) into a valine (APP[E691V]), did not modify ADAM30 wild-type phenotype. These results thus indicate that only a mutation within the lysosome addressing sequence (APP[F690S]) abolished ADAM30-mediated impact on APP metabolism. Of note, we did not detect any modification on APP[WT] endocytosis in line with ADAM30[WT] over-expression (data not shown).

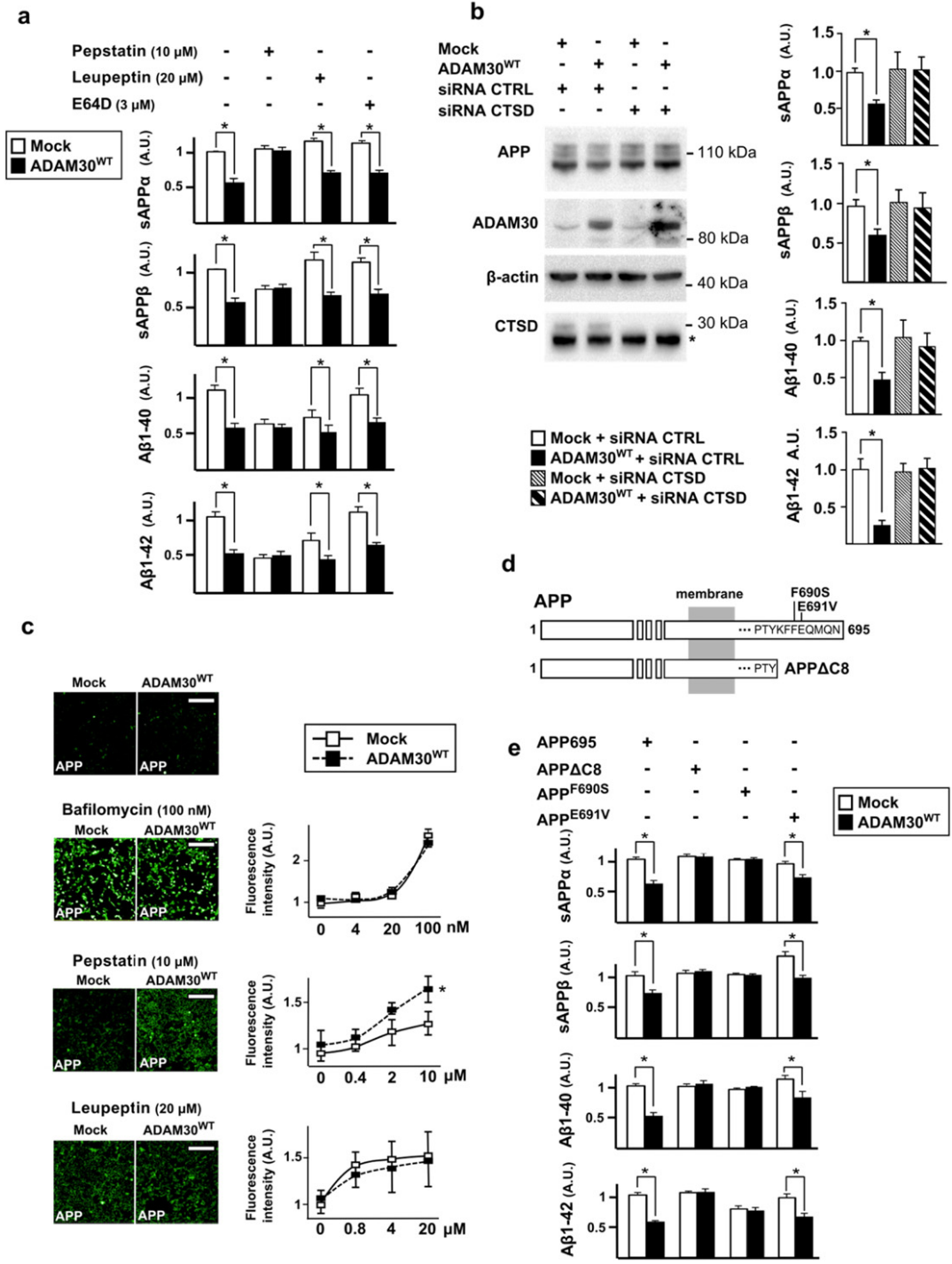

**Fig. 4.** Assessment of the impact of ADAM30 on the APP metabolism through CTSD activation. (a) Impact of pepstatin, leupeptin or E64D treatment in HEK293-APP695$^{WT}$ cells 48 h after transfection with ADAM30$^{WT}$ or mock vector. Cells were treated for 24 h at the indicated concentration. Mean differences (±SEM) in the amounts of sAPPα, sAPPβ, Aβ1–40 and A-1β42 are shown. (b) Impact of CTSD inhibition by siRNA on APP metabolism according to ADAM30$^{WT}$ over-expression. Mean differences (±SEM) in the amounts of sAPPα, sAPPβ, Aβ1–40 and A1β-42 are shown. (c) The impact of inhibiting APP degradation in HEK293-APP695$^{WT}$-ADAM30$^{WT}$ cells. APP degradation was blocked by 24 h of treatment with bafilomycin, pepstatin or leupeptin at the indicated concentration. Intracellular APP accumulation was assessed using immunofluorescence staining with LN27 antibody against the N-terminal part of APP. Scale bar: 100 μm. The graph represents the mean difference (±SEM) in fluorescence intensity per cell (n > 150) for the indicated concentration. (d) Schematic representation of the different mutation used in the C-ter APP. (e) The impact of ADAM30 overexpression on the metabolism of C-ter-mutated APP in HEK293 cells. Mean differences (±SEM) in the secreted amounts of sAPPα, sAPPβ, Aβ1–40, and Aβ1–42 are shown. In all panels, three independent experiments were performed in duplicate after 24 h of transfection. *$p < 0.05$ (Mann-Whitney non-parametric test).

### 4.8. Overexpression of ADAM30 Modulates Aβ Production and Deposition In Vivo

To corroborate our observations in vivo and further evaluate the impact of ADAM30 on Aβ peptide secretion and amyloidosis, we sought to generate a transgenic mouse model. It was not possible to develop ADAM30 knock-out mice (ADAM30 is not expressed in the mouse brain (Allen Brain Atlas, http://www.brain-map.org and our data, Fig. 5a and Supplementary Fig. 13) contrary to what we observed in humans. We thus generated APP[Sw,Ind] mice that conditionally overexpressed hADAM30[WT] in the neurons of the forebrain (as previously mentioned, immunohistochemistry experiments were consistent with the expression of human ADAM30 in neurons in human brain tissue (Fig. 1c)). Primary cultures of adult neurons indicated that expression of ADAM30[WT] was only observed in neurons generated from hADAM30[WT]-hAPP[Sw,Ind]-Cre mice (Fig. 5a). As observed in cell lines, a decrease in APP catabolites was also observed in primary cultures of adult neurons expressing ADAM30[WT] (Fig. 5b).

We then extended our results by analyzing the level of soluble Aβ42 in the mouse hippocampus. We observed significantly lower hippocampal levels of soluble Aβ42 upon induction in mice overexpressing hADAM30[WT] ($-39\%$, relative to hADAM30[WT]-hAPP[Sw,Ind] mice; $p = 0.03$). We also generated mice that conditionally overexpressed hADAM30[mut] in order to confirm that ADAM30-mediated potential effects are genuinely linked to its catalytic activity. Remarkably, we did not detect any lower levels of soluble Aβ42 in these mice overexpressing hADAM30[mut] (Fig. 5c). CTSD activity was also significantly higher in the cortex of mice overexpressing hADAM30[WT] (but not in those overexpressing hADAM30[mut]) than in hADAM30-hAPP[Sw,Ind] mice ($+34\%$, $p < 0.0001$, and $+7\%$, $p = 0.82$, respectively; Fig. 5D). Interestingly, we also observed a negative correlation between CTSD activity and soluble Aβ42 concentrations in hADAM30[WT]-hAPP[Sw,Ind]-Cre mice ($p < 0.05$ in Spearman's correlation test, Fig. S10).

All these observations are in full accordance with our in vitro data and support the claim that our observed effects are indeed specifically due to ADAM30 enzymatic activity. Furthermore, our results support a direct link between ADAM30, CTSD activation and amyloid load.

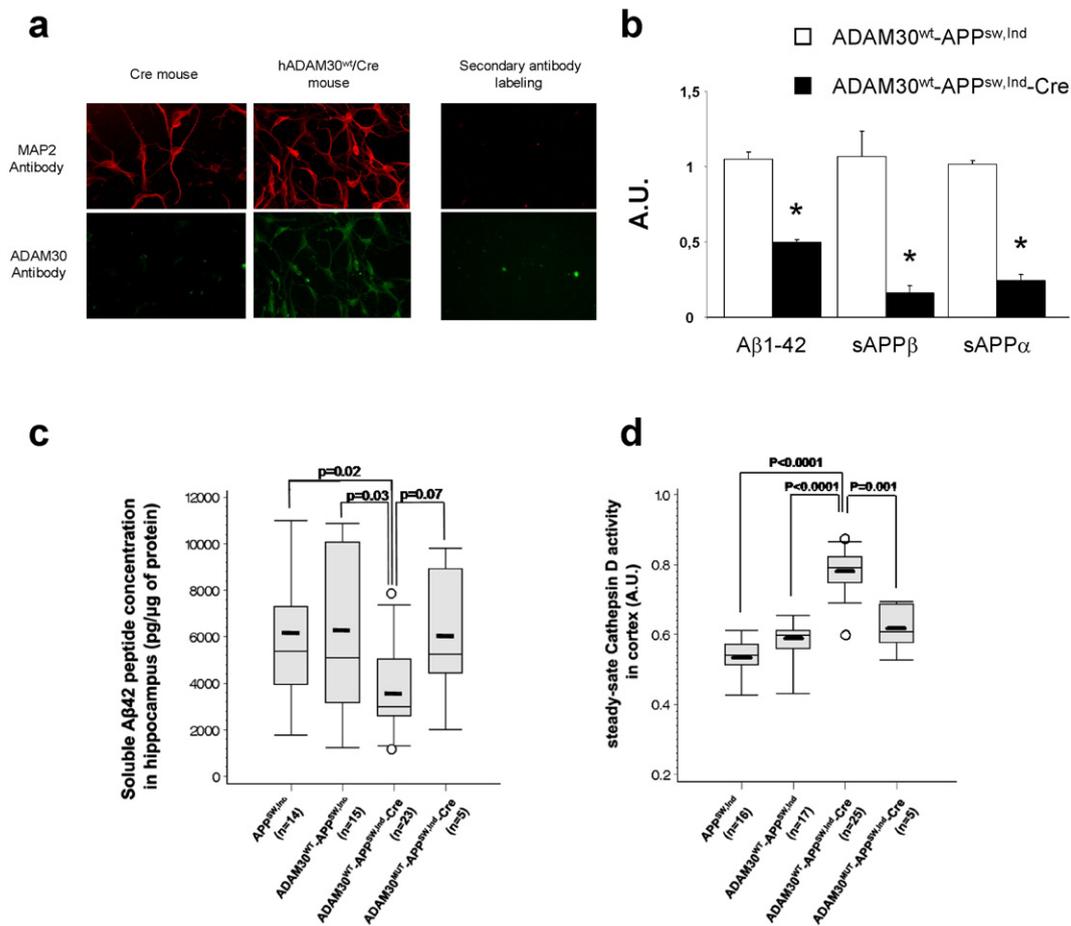

**Fig. 5.** Impact of ADAM30 over-expression in neurons and AD-like models on Aβ secretion and CTSD activity. (a) ADAM30 expression at the protein level in adult neurons from 2 month mouse hippocampus ($n = 3$). (b) Mean differences ($\pm$SEM) in the secreted amounts of sAPPα, sAPPβ, and Aβ1–42 in adult neurons (Div 19) according to ADAM30[WT] over-expression. Three independent experiments were performed in duplicate. *$p < 0.05$ (Mann-Whitney non-parametric test). (c) Differences in levels of soluble Aβ1–42 peptide concentration in the hippocampi of different transgenic mice overexpressing ADAM30[WT/mut] or not. The thick lines represent the median Aβ1–42 level. The midline represents the mean value and the upper and lower horizontal lines represent the first and third quartiles, respectively. Circles indicate individuals with extreme values (more than 2 SD above or below the mean value). p-Values refer to a Mann-Whitney non-parametric test. (d) Differences in CTSD activity in the cortex of different transgenic mice overexpressing ADAM30[WT/mut] or not. The thick lines represent the median CTSD activity level. The midline represents the mean value and the upper and lower horizontal lines represent the first and third quartiles, respectively. Circles indicate individuals with extreme values (more than 2 SD above or below the mean value). p-Values refer to a Mann-Whitney non-parametric test.

To assess this possibility, we used a reliable, automated, high-throughput, three-dimensional (3D) histology method to determine the Aβ load in mouse forebrain in general and the hippocampus in particular (see the Experimental procedures, Fig. 6a). Current standards for the analysis of brain histopathological markers heavily rely on manual intervention to delineate regions of interest and quantify the staining. Data collection is thus usually restricted to a few tissue sections. As a consequence, this approach allows avoiding eventual biases due to low section sampling rate as demonstrated elsewhere (Vandenberghe et al., 2016). By using this powerful approach, we observed that overexpression of hADAM30$^{WT}$ was associated with a 20% reduction in the Aβ load ($p = 0.03$; Fig. 6b) in the mice forebrain. As expected, this reduction was also observed in analysis restricted to the hippocampus ($-28\%$, $p = 0.01$; Fig. 6c).

### 4.9. Overexpression of ADAM30 Partially Rescues LTP Deficits

Since APP catabolites have been shown to block hippocampal long term potentiation (LTP) and impair learning function in mice (Nalbantoglu et al., 1997), we aimed at establishing the functional impact of hADAM30 overexpression on LTP in hAPP$^{Sw,Ind}$ hippocampus. Several paradigms of basal synaptic transmission were analyzed: maximum fEPSP slope, Maximum fEPSP amplitude, maximum fiber volley amplitude and the function relating the stimulation intensity to fEPSP, which is an index of basal synaptic strength. None of these parameters were altered whatever the transgenic mice analyzed (Fig. 7a).

As expected, a LTP deficit was observed in hAPP$^{Sw,Ind}$ and hADAM30$^{WT}$-hAPP$^{Sw,Ind}$ mice and this deficit was partially restored over time in hADAM30$^{WT}$-hAPP$^{Sw,Ind}$-Cre mice (Fig. 7a and b). Mean fEPSP for the different groups was calculated over the last 20 min of recording post-high frequency stimulation. A significant difference in LTP was observed between the hADAM30^WT-hAPP^Sw,Ind-Cre mice and the other transgenic mice in which no potentiation was observed. Even if ADAM30^WT over-expression did not fully rescue potentiation, an about 60% recovery was observed (Fig. 7c).

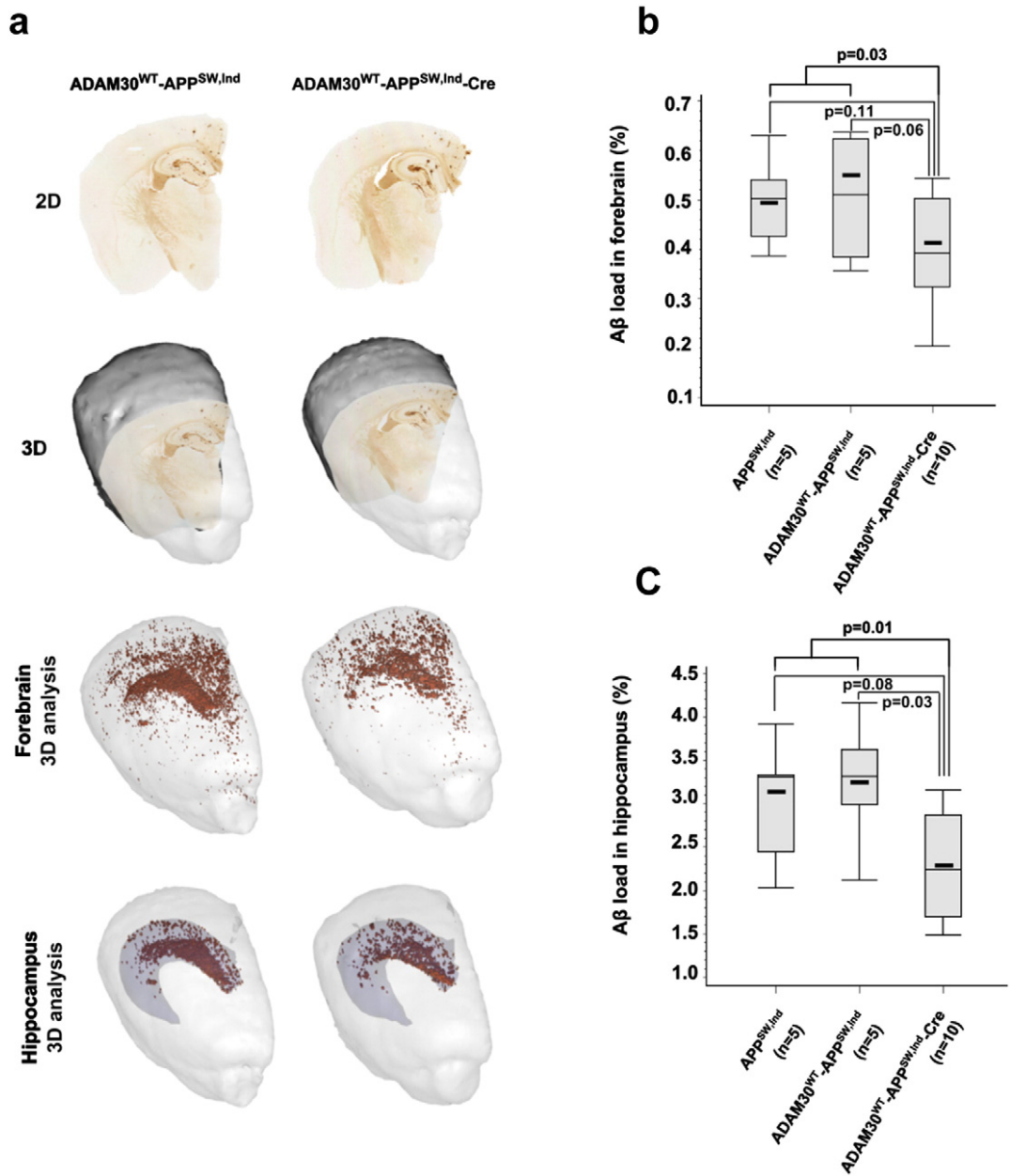

**Fig. 6.** Impact of ADAM30 over-expression in AD-like models on Aβ loads. (a) Representative 2D sections and 3D reconstruction of the brains in mice overexpressing ADAM30$^{WT}$ or not. Illustration in the same mice of the spatial distribution of 3D segmented Aβ peptide aggregates corresponding to forebrain and hippocampus regions (b) The Aβ load in the forebrain of APP$^{sw,Ind}$ transgenic mice overexpressing ADAM30$^{WT}$ or not. (c) The Aβ load in the hippocampus of APP$^{sw,Ind}$ transgenic mice overexpressing ADAM30$^{WT}$ or not. Box plots are as follows: The thick lines represent the median level. The midline represents the mean value and the upper and lower horizontal lines represent the first and third quartiles, respectively. Circles indicate individuals with extreme values (more than 2 SD above or below the mean value). p-Values refer to a Mann-Whitney non-parametric test.

## 5. Discussion

One of the main difficulties in differential expression analyses of postmortem tissues relates to whether the variations detected are causes or consequences of the disease process. By means of a multidisciplinary approach involving various state-of-the-art techniques and novel original transgenic mouse models, we identified ADAM30 as a player in APP metabolism and show that its underexpression likely directly contributes to Alzheimer's disease physiopathology.

Very little is known about the physiological role of ADAM30 (Cerretti et al., 1999; Hu et al., 2009; Ho et al., 2013; Almawi et al., 2013; Gupta et al., 2012; Ellis et al., 2014), its putative involvement in Alzheimer's disease and its protein partners. However, our results highlight a new axis in APP metabolism, with APP sorting to lysosomes, ADAM30-dependent CTSD activation and APP degradation. Overall, this set of data supports previous mechanistically unsolved observations of an important lysosomal contribution to the metabolism of APP (Funk and Kuret, 2012).

The potential APP degradation through ADAM30-dependent CTSD activation could be either linked to the CTSD activity by itself (CTSD was described to endoproteolyse the C100 APP fragment at multiple sites and to display β-secretase-like activity, in vitro) (Mackay et al., 1997; Chevallier et al., 1997) or to a CTSD-mediated subsequent activation of other hydrolases (Benes et al., 2008). In line with the recent description that PS1 mutations can be linked to defective lysosomal proteolysis (Lee et al., 2010), our data support the importance of the lysosome compartment in APP metabolism and as a consequence in the Alzheimer's disease pathophysiological processes.

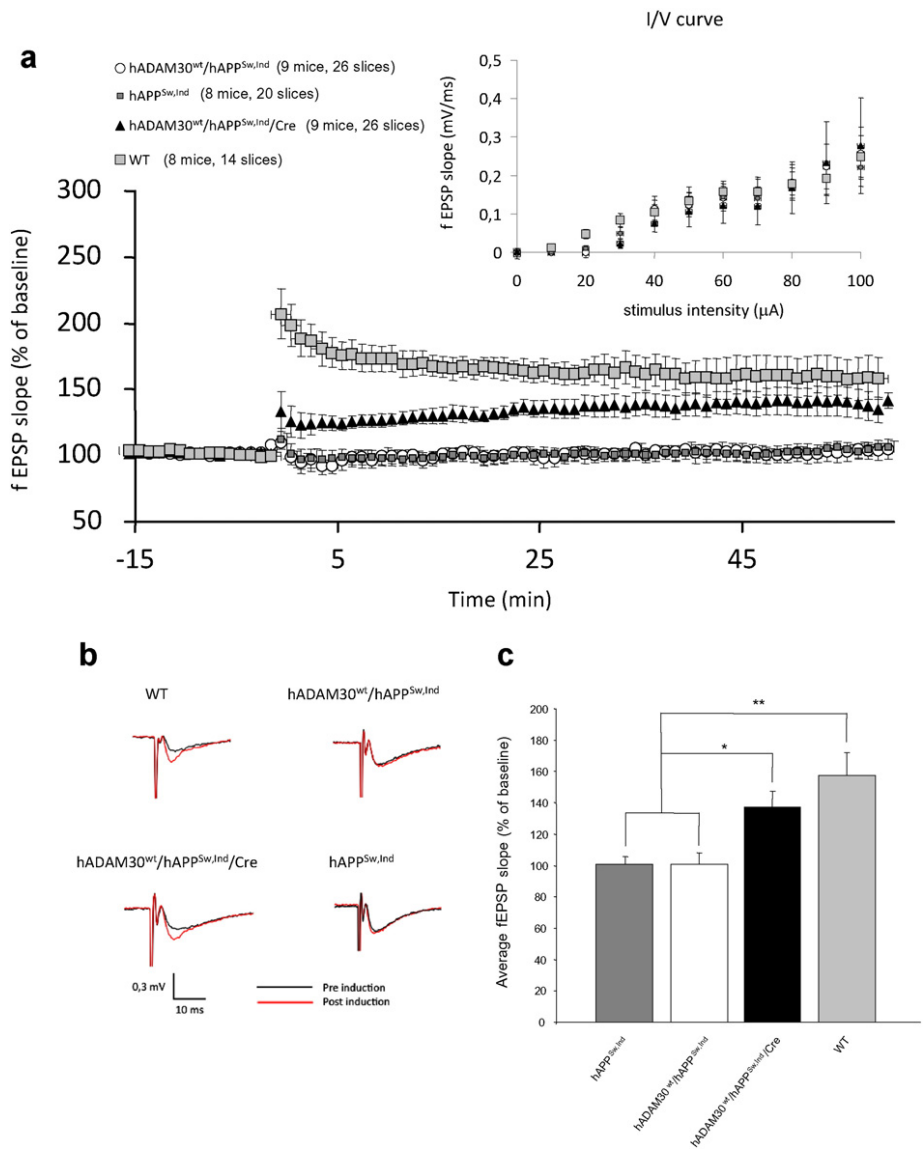

**Fig. 7.** Impact of ADAM30^WT over-expression on LTP in Alzheimer-like mouse model. (a) LTP at the CA3-CA1 synapses of hippocampal slices induced by $4 \times 100$-Hz stimulation is impaired in hADAM30^WT-hAPP^Sw,Ind (nine mice) and hAPP^Sw,Ind (eight mice) compared with WT mice (eight mice). Cre dependent gene expression of hADAM30^WT in hADAM30^WT-hAPP^Sw,Ind-Cre mice partially rescues LTP (nine mice). Inset graph shows input–output curves, which are comparable in hADAM30^WT-hAPP^Sw,Ind (nine mice), hAPP^Sw,Ind (n = 8 mice), hADAM30^WT-hAPP^Sw,Ind-Cre (n = 9 mice), and WT mice (eight mice). (b) Traces show the field excitatory postsynaptic potential (fEPSP) from a representative animal from each group, black traces represent the average of baseline recording before $4 \times 100$-Hz stimulation, red traces show the average over the last 20 min of post-stimulation recording. (c) Quantification indicating the percentage change in average fEPSP amplitude 40–60 min after $4 \times 100$-Hz stimulation. Data are mean ± SEM. **$p = 0.0005$, **$p = 0.0015$ (Mann-Whitney non-parametric test).

We suspect that further investigation of other potential ADAM30 substrates might provide a better understanding of the ADAM30-dependent physiopathologic processes. Interestingly, all these targets can be directly or indirectly linked to insulino-resistance and/or APP metabolism. In one hand, GKAP1 was recently described to interact with IRS1 and to participate in insulino-resistance in adypocites (Ando et al., 2015) whereas the ADAM30 locus was reported as a genetic risk factor for Type 2 diabetes (T2D) (Cerretti et al., 1999; Hu et al., 2009; Ho et al., 2013; Almawi et al., 2013; Gupta et al., 2012; Ellis et al., 2014). On the other hand, IRs, insulino-resistance and TD2 were described as modulators of the Alzheimer's disease pathology though a potential modulation of the APP metabolism whereas CTSD was previously involved in APP processing in vitro (Dreyer et al., 1994; Mackay et al., 1997; Haque et al., 2008; Funk and Kuret, 2012). It is also noteworthy that IRS4 (also identified in our COFRADIC study) is able to directly interact with the Grb2 adaptor protein, a well-known actor of APP processing/sorting (Hinsby et al., 2004). Of note, since lysosomal proteolysis can represent a general mechanism for protein degradation/processing, we measured secreted catabolites of Met, a protein exhibiting a metabolism similar of APP (Lefebvre et al., 2012). We did not observe any modifications of the secreted Met products following co-transfection of Met and ADAM30 in HEK293 cells (Fig. S11). This observation suggests that the ADAM30-dependent CTSD-lysosome activation may be rather selective and that its metabolic consequences could be limited to a restricted number of proteins.

When considering the APP metabolism as a whole, our present findings (including the ADAM30-linked decreased secretions of APPs-α and -β and Aβ peptides) strongly suggest that either (i) ADAM30 is involved in a process that modifies the equilibrium between APP recycling (from early/late lysosomes to the membrane) and APP degradation in the lysosome or (ii) ADAM30 affects the cellular events that drives the dispatching of full-length membrane APP to either the endocytosis pathway or the lysosomal compartment, as was previously suggested (Lorenzen et al., 2010). However, it is noteworthy that we did not detect any changes in the internalization or endocytosis of full-length membrane APP in cells overexpressing ADAM30.

In conclusion, we hypothesize that ADAM30 is a key player in APP metabolism; it appears to be involved in APP degradation via sorting to lysosomes, targeted CTSD activation and then full-length APP recycling. The mechanisms that involve ADAM30 may restrict full-length APP recycling and therefore influence the pool of APP available for α-, β- and γ-secretase processing.

In conclusion, we characterized a new facet of the APP physiology which may provide new therapeutic targets to control amyloid peptide production in the brain.

Supplementary data to this article can be found online at http://dx.doi.org/10.1016/j.ebiom.2016.06.002.

## Conflict of Interest

The authors declare no conflict of interest.

## Author contributions

F. Letronne., G.L., A-M.A., J.C., A.F, F.D, F.E and T.G. performed molecular and cellular biology experiments. G.L., A-M.A., J.C., F.D., N.M., E.W. and F. Lafont performed/supervised cellular imaging. F. Letronne., A-M.A., Y.S. performed mouse model breeding, brain preparation and soluble Aβ measurements. G.L., A.B., F.H. performed Quantigene assays. M.L., K.G. performed COFRADIC experiments and interpretation. M.E.V, A-S.H., M.D., N.S. and T.D. performed 3D brain amyloid load analyses and interpretation. F. Leroux., J.D., B.D., R.D. developed and performed recombinant CTSD/ADAM30 assays. M.L. and D.T. performed Met experiments. L.H., Y.L., D.H. performed microarray experiments and interpretation. L.C., C.B., F.C. performed β- and γ-secretase assays. F.P., C.B., J-J.H., D.M. provided human brain samples for immunohistochemistry and transcriptomic analyses. A.-M.A., J.C., M.E.V, A.B., P.D., C.D., B.D., F. Lafont, M.D., D.M., P.A., F.C., D.H., T.D., J-C.L. participate to the writing and/or significantly revise the manuscript. A-M.A, J.C. M.L., R.D., D.H., F.C., T.D., J-C.L. conceived the experiments.

## Acknowledgments

We thank the Lille Neurobank and UMR 1172 "Alzheimer & Tauopathies" for the brain immunohistochemistry experiments. We thank Florence Combes for her technical assistance. We thank Dr. Melissa Farinelli (E-Phy-Science) for her helpful discussion. F.L. was funded by the University of Lille II, the Nord-Pas de Calais Regional Council and the Fondation pour la Recherche Médicale (FRM). G.L. and F.E. was funded by the Institute Pasteur de Lille and the Nord-Pas de Calais Regional Council. J.C. and Y.S. were funded by the MEDIALZ Project (Grant 11001003) financed by the European Regional Development Fund and the Nord-Pas de Calais Regional Council. F.H. was funded by the Alzheimer's Association (grant IIRG-06-25487) and then the Ligue contre la Maladie d'Alzheimer (LECMA, grant 09705). C.D. was funded by Lille Metropole Communauté Urbaine. F.C. was supported by the Hospital University Federation (FHU OncoAge).

This work was also funded by the French government's CPER-Neuroscience (DN2M) program (Nord-Pas de Calais Regional Council and FEDER), INSERM (the ATC-vieillissement program), the Institut Pasteur de Lille, the Fondation pour la Recherche sur le Cerveau (FRC) and the French government's "Development of Innovative Strategies for a Transdisciplinary approach to Alzheimer's disease" Laboratory of Excellence (LABEX DISTALZ) program.

We acknowledge the support of Alzheimer's Research UK and Alzheimer's Society through their funding of the Manchester Brain Bank (from which samples were also obtained) under the Brains for Dementia Research (BDR) initiative.